**A Structure-Preserving Assessment of VBPBB for Time Series Imputation Under Periodic Trends, Noise, and Missingness Mechanisms**

Asmaa Ahmad[1]*, Eric J Rose[1], Michael Roy[2], Edward Valachovic[1]

[1] Department of Epidemiology and Biostatistics, College of Integrated Health Sciences, University at Albany, State University of New York, One University Place, Rensselaer, NY

[2] NYS Department of Health

*Corresponding author.

E-mail: Aahmad4@albany.edu (AA)

## Abstract

Incomplete time-series data compromise statistical inference, particularly when the underlying process exhibits periodic structure (e.g., annual or monthly cycles). Conventional imputation procedures rarely account for such temporal dependence, leading to attenuation of seasonal signals and biased estimates. This study proposes and evaluates a structure-preserving multiple imputation framework that augments imputation models with frequency-specific covariates derived via the Variable Bandpass Periodic Block Bootstrap (VBPBB)**.** In controlled simulations, we generate series with annual and monthly components, impose Gaussian noise across low, moderate, and high signal-to-noise regimes, and introduce Missing Completely at Random (MCAR) patterns from 5% to 70% missingness. Dominant periodic components are extracted with VBPBB, resampled to stabilize uncertainty, and incorporated as covariates in Amelia II. Compared with baseline methods that do not model temporal structure, the VBPBB-enhanced approach consistently yields lower imputation error and superior retention of periodic features, with the largest gains observed under high noise and when multiple components are included. These findings demonstrate that explicitly modeling periodic content during imputation improves reconstruction accuracy and preserves time-series structure in the presence of substantial missingness.

### Policy Significance Statement

Missing data is a major challenge in public health monitoring, reducing the reliability of reporting systems and weakening evidence used for planning, surveillance, and resource allocation. This study introduces an imputation framework that incorporates periodic signals, such as seasonal and monthly patterns, to more accurately reconstruct missing values in time series data. The approach demonstrates substantial accuracy gains across multiple noise and missingness conditions compared to standard methods. By preserving underlying temporal structure, this technique improves data quality and strengthens confidence in population-level trends. Policymakers and analysts can apply these findings to generate more reliable estimates from incomplete datasets, support more informed decision-making, and reduce uncertainty in systems that rely on continuous reporting.

## Overview

The study evaluates whether incorporating frequency-targeted, bootstrapped periodic components into a multiple-imputation workflow improves accuracy and preserves temporal structure compared with methods that do not model temporal dependence. We develop a structure-preserving framework in which the Variable Bandpass Periodic Block Bootstrap (VBPBB) is used to isolate statistically significant periodic components, principally annual and monthly signals, with additional harmonics included when warranted, and to generate stabilized, band-limited covariates. These covariates are then incorporated into Amelia II to perform multiple imputation on incomplete time-series data, and performance is benchmarked against comparison models that exclude periodic covariates.

To assess performance under controlled conditions, we simulate series that embed annual and monthly periodicities and add Gaussian noise spanning low, moderate, and high signal-to-noise regimes. Missingness is imposed under a Missing Completely at Random (MCAR) mechanism at rates from 5% to 70%. Frequency selection relies on significance screening of candidate bands, after which VBPBB provides resampled component summaries that enter the imputation model as auxiliary variables. Evaluation focuses on reconstruction accuracy using root mean squared error (RMSE) and mean absolute error (MAE), complemented by measures of structure preservation such as band-limited power retention and fidelity of seasonal amplitude and phase.

Across all simulated conditions, the VBPBB-enhanced approach delivers lower imputation error and superior retention of periodic features relative to the comparison methods. Gains are most pronounced in high-noise scenarios and when multiple periodic components are supplied to the imputation model, indicating that explicitly modeling frequency-specific structure mitigates noise-induced attenuation and better reconstructs seasonal dynamics.

The principal contributions are a general, structure-preserving imputation framework for periodic time series, a simulation protocol that spans realistic ranges of noise and missingness, and practical guidance for frequency selection and component inclusion. Current experiments emphasize MCAR settings; extending the framework to MAR/MNAR mechanisms, nonstationary regimes, and formal assessments of interval coverage and downstream inferential validity are important directions for future work.

## 1. Background

Missing data are common in applied research and pose substantive threats to statistical inference and machine-learning pipelines. Practitioners often resort to ad hoc or convenience methods to fill gaps, yet the choice of imputation strategy is consequential: it conditions the quality of the completed data and can materially affect the accuracy, stability, and interpretability of subsequent models (McNeish, 2016; van Buuren, 2018). Consequently, handling missingness is not a peripheral preprocessing step but a central determinant of model validity.

Time-series settings are especially vulnerable to missingness arising from equipment failures, reporting delays, and other forms of longitudinal nonresponse (Schafer & Graham, 2002). Such data frequently exhibit periodic structure, annual, monthly, or weekly cycles, that is integral to correct interpretation. Commonly used approaches (e.g., mean substitution, linear interpolation, and off-the-shelf multiple imputation) typically ignore serial dependence and seasonal structure, which can attenuate amplitudes, distort phase, and degrade forecasting and inference (Ahn et al., 2020; Zhang, 2016; Little & Rubin, 2002). The result is imputed values that fail to reflect the true dynamics of the process and bias that propagates into downstream analyses.

Preserving temporal structure during imputation is therefore essential in domains where time-dependent patterns guide interpretation and policy, public health surveillance, systems monitoring, and environmental risk assessment among them. Effective strategies must leverage information in the series itself to maintain continuity and coherence and to reduce bias in estimates derived from completed datasets (Zhang, 2016).

To address these limitations, we develop a structure-preserving imputation framework that explicitly incorporates periodic components using the Variable Bandpass Periodic Block Bootstrap (VBPBB) (Valachovic, 2024). VBPBB isolates statistically significant frequency components while resampling in a manner that preserves temporal autocorrelation and phase relationships. The resulting band-limited, bootstrapped components are then introduced as auxiliary covariates within the Amelia II multiple-imputation model, enabling reconstruction of missing values in a way that retains the series' underlying frequency structure.

## 2. Overview of the Imputation Process

Figure 1 presents the end-to-end pipeline for imputing incomplete time-series data. We begin with an observed series containing missing values and first recover dominant periodic structure using the Variable Bandpass Periodic Block Bootstrap (VBPBB) (Sec. 3.4). For each retained frequency band, VBPBB generates bootstrap replicates that preserve temporal dependence by resampling in period-matched blocks. We summarize these replicates by their frequency-wise median, yielding stable, band-limited components that capture the principal seasonal signals while attenuating off-target noise.

These frequency-specific components are then incorporated as auxiliary covariates in the multiple-imputation model implemented in Amelia II (Sec. 3.5). Let B denote the number of VBPBB bootstrap replications and m the number of completed datasets produced by Amelia per run. In our primary specification we set $m=1$ and treat the B runs as the multiple imputations; in robustness checks we allow $m>1$ and combine all $B \times m$ completions. Within each run, Amelia's expectation–maximization (EM) algorithm and data augmentation draw on the periodic covariates to reconstruct missing entries while preserving the estimated frequency structure.

The resulting completed datasets are then pooled to obtain point estimates and uncertainty following standard multiple-imputation rules (Sec. 3.6). Optionally, we apply light post-imputation smoothing (e.g., a low-order Kolmogorov–Zurbenko filter or short moving average) to improve local coherence without blurring the recovered seasonal amplitudes (Sec. 3.5). Performance is evaluated against the available ground truth using mean absolute error (MAE) and root mean squared error (RMSE), with supplementary diagnostics targeting structure preservation (e.g., retention of band-limited power and seasonal phase alignment) reported where relevant (Sec. 3.6).

Finally, all hyperparameters are specified a priori: the VBPBB passbands and block lengths (matched to the targeted periods), the number of bootstrap replications B, the set of auxiliary covariates (including optional lags), Amelia's m and priors, and the smoothing kernel (if used). Section 3 details frequency selection, bootstrap design, covariate construction, and pooling

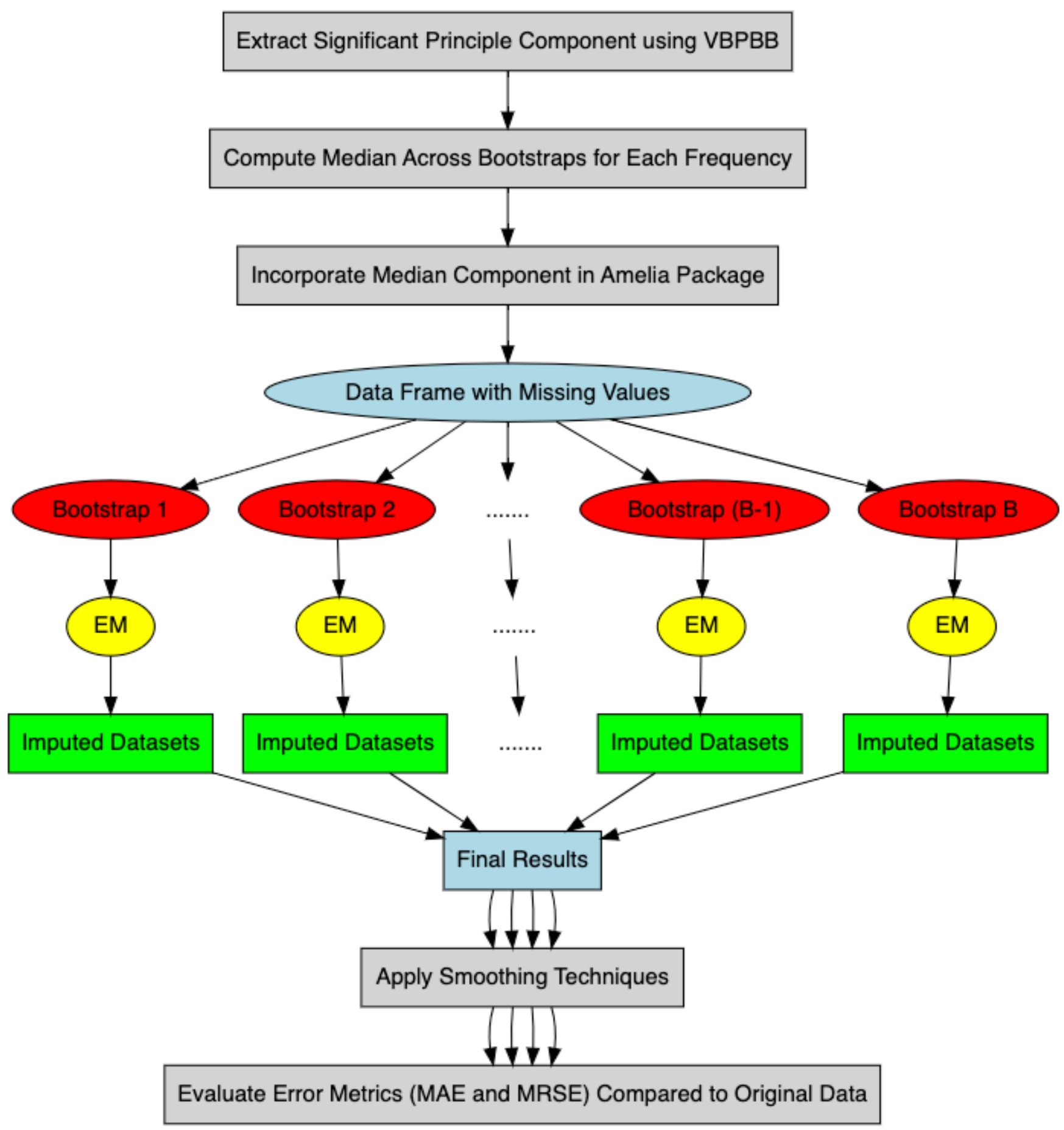

**Figure 1:** Flowchart of Data Imputation Process Using VBPBB

# 3. Method

## 3.1. Data Generation and Simulation Setup

To rigorously assess the proposed imputation methodology under controlled yet realistic conditions, we implemented a tiered simulation using synthetic daily time-series data. Each dataset comprised 6,000 observations (≈16.4 years) generated by superimposing known periodic signals with additive Gaussian noise. The base signal combined sinusoidal components chosen to mimic common seasonality: an annual cycle (frequency = 1/365), its first harmonic (2/365), and a monthly cycle (1/30), depending on the experimental stage; all component amplitudes were fixed at 10. Noise variance was varied across $\sigma^2 \in \{0.1, 4, 10, 25, 100\}$ to span low through very high signal-to-noise scenarios and test robustness. Prior to frequency extraction, we removed a smooth trend (local regression, span = 0.2) and standardized the residual series; imputations were back-transformed to the original scale. The overall workflow these simulations feed into is summarized in the imputation pipeline (see Figure 1), while representative 600-day segments illustrate how increasing noise progressively masks these structures across the three signal stages (Figure 2).

The simulation study was structured across three levels of increasing signal complexity:

1. **Stage 1: Single Annual Cycle**
   Data were generated using only the annual periodic component. Each dataset was then subjected to five levels of noise and five missing data proportions (5%, 10%, 15%, 20%, and 70%). Both baseline and VBPBB-enhanced imputation methods were applied to evaluate reconstruction accuracy.
2. **Stage 2: Annual Cycle + First Harmonic**
   The signal from Stage 1 was extended by adding the first harmonic. The same noise and missingness conditions were applied, and both baseline and VBPBB-enhanced imputation methods were used to assess the impact of added signal complexity on reconstruction accuracy.
3. **Stage 3: Multi-Component Signal (Annual + Harmonic + Monthly)**
   A monthly component (1/30) was incorporated to create a composite signal reflecting more intricate periodically correlated dynamic. As before, the datasets underwent identical noise and missingness conditions and were imputed using both baseline and VBPBB-enhanced methods.

Each configuration was replicated multiple times (30 replication) with fixed random seeds to ensure statistical reliability and reproducibility. The original complete series served as the reference for evaluation. We report **MAE** for interpretability and **RMSE** to emphasize large deviations, using the definitions provided in Chapter 3.

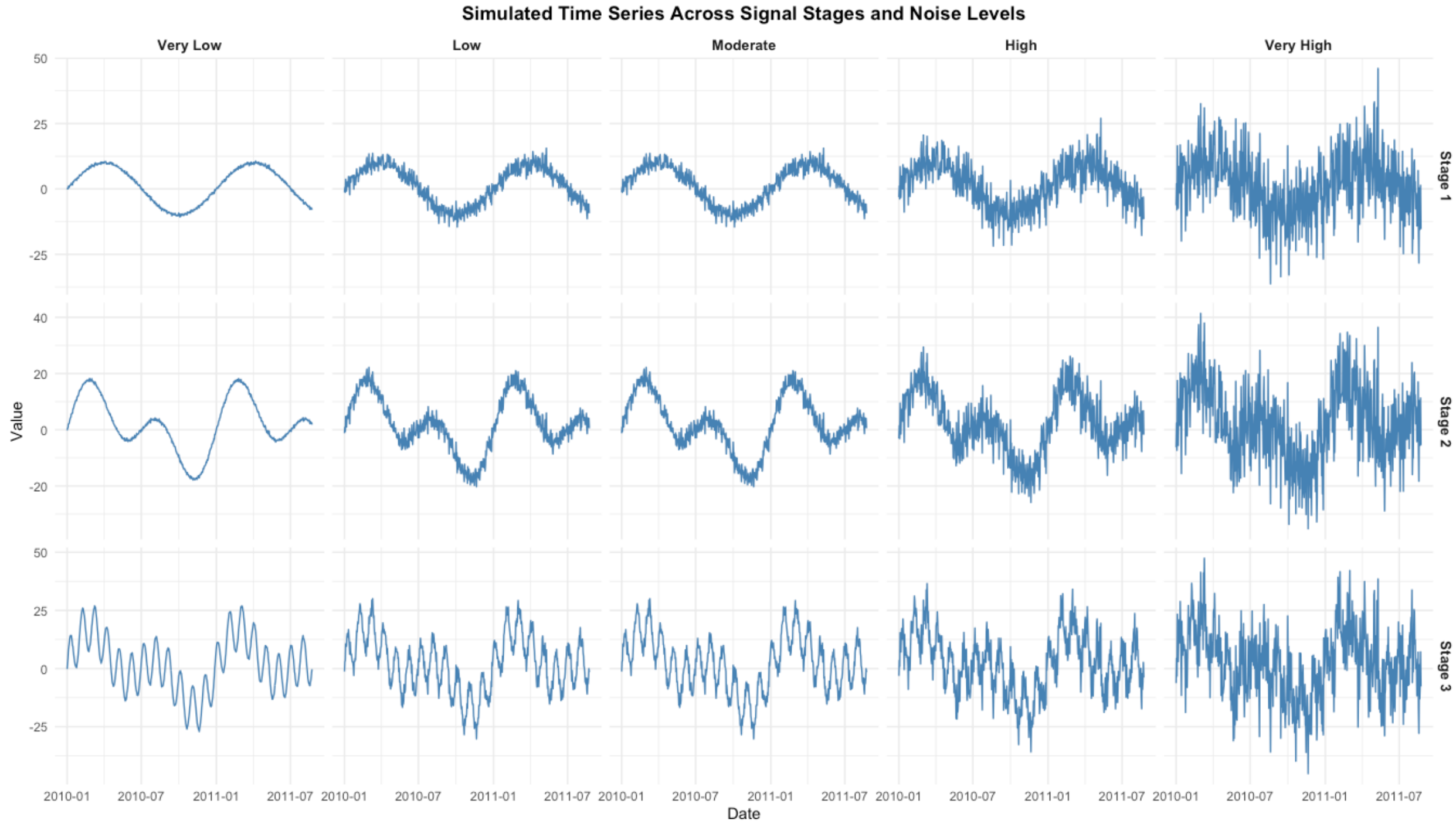

**Figure 2.** Simulated time series across signal stages and noise levels.

Figure 2. Simulated time-series panels by signal complexity and noise level. Each panel depicts a 600-day segment of the synthetic data used in the experiments, arranged by signal stage (rows) and noise variance (columns). Stage 1 includes only the annual component (frequency 1/365); Stage 2 adds the first harmonic (2/365); Stage 3 adds the monthly component (1/30). Noise increases left-to-right from very low ($\sigma^2$ =0.1) to very high ($\sigma^2$ =100). Panels illustrate how periodic structure is increasingly obscured as noise grows, contextualizing the imputation task complexity shown in Figure 1. Missingness Completely at Random (MCAR) is introduced later and is not depicted here; see Section 3.2 ("Missing Data Mechanism").

### 3.2. Missing Data Mechanism

Missing values were introduced using a **Missing Completely at Random (MCAR)** mechanism. Under this assumption, the probability of a value being missing is independent of both observed and unobserved data. This approach reflects an idealized scenario in which data are missing purely by chance, without any systematic bias or underlying pattern.

Missing values were introduced exclusively in the primary time series variable to focus the evaluation on the imputation method's capacity to reconstruct the principal signal. The auxiliary covariates, specifically, the periodic components extracted through the VBPBB framework (annual, harmonic, and monthly), were retained as complete to ensure they consistently provided supplementary information across all missing data scenarios (see Section 3.5.2 Model Covariates for a detailed description).

For each simulation run, missingness was applied by randomly selecting a fixed proportion of observations from the full time series. Specifically, we generated incomplete datasets by removing **5 %, 10 %, 15** %, **20 %,** or **70 % of** the total data points using uniform sampling without replacement. Missing values were introduced exclusively in the primary time series variable to focus the evaluation on the imputation method's capacity to reconstruct the principal signal. The auxiliary covariates were retained as complete to ensure they consistently provided supplementary information across all missing data scenarios. This approach isolates the impact of missingness on the target variable and avoids confounding effects that could arise from simultaneously introducing missing data into the covariates.

The MCAR framework allows for a controlled comparison of imputation strategies by isolating the effects of missingness from potential confounding due to data structure or distribution. Each missingness level was tested across all noise conditions and signal complexity stages, with repeated simulations conducted to ensure the robustness and reproducibility of results.

### 3.3. Noise Level Selection

We examined performance under varying signal quality by adding independent, mean-zero Gaussian noise at five variances **0.1 (very low noise)**, **4 (low noise)**, **10 (moderate noise)**, **25 (high noise)**, and **100 (very high noise)**. These levels approximate a realistic spectrum of signal-to-noise ratios encountered in longitudinal public-health and environmental monitoring applications. As noise increases, periodic structure becomes less distinguishable, posing a progressively more stringent test of each imputation strategy's ability to recover the underlying pattern and preserve seasonal properties.

### 3.4. Periodic Component Extraction Using VBPBB

To improve the structural accuracy of imputed values, we extracted dominant periodic components from the fully observed time series using the Variable Bandpass Periodic Block Bootstrap (VBPBB) framework. This approach preserves both short- and long-range autocorrelation structures that are often distorted by conventional imputation methods, particularly in noisy or seasonal time series. The VBPBB method is specifically designed to apply filters to principal component (PC) time series data, segmenting and isolating distinct frequency components through bandpass filters centered between target frequencies (Valachovic and Shishova, 2025; Valachovic, 2025). Its implementation incorporates the Kolmogorov–Zurbenko Fourier Transform (KZFT), a nontraditional spectral technique that replaces classical Fourier transforms with moving averages and convolution operators in the time domain (Zurbenko, 1986). For each simulation, we targeted three predefined frequencies: annual (1/365), semiannual (2/365), and monthly (1/30), selected for their relevance to real-world seasonal and administrative cycles. The annual and semiannual components reflect broad seasonal patterns, such as flu seasonality or cyclical hospitalization trends (CDC, 2025; Zurbenko & Cyr, 2011), while the monthly component captures localized operational rhythms, such as billing or reporting cycles (Austin et al., 2021). This choice of fundamental and harmonic frequencies mirrors the structure observed in many real-world datasets beyond healthcare, where multiple periodic components and their harmonics coexist and interact (Valachovic, 2024; Yang & Zurbenko, 2010). Thus, our framework is adaptable to contexts where distinct temporal patterns, whether quarterly, weekly, or even daily, inform predictive modeling or imputation strategies.

The KZFT was used to decompose the signal through a series of weighted moving averages that iteratively smooth the data, extracting frequency-specific content at multiple scales. Key parameters in this filtering process include the frequency center (ν), defined as the reciprocal of the period (1/p), which ensures symmetric distribution around the target frequency (Ahmad et al., 2025). The filter's bandwidth was controlled through careful tuning of the moving average window width (m) and the number of iterations (k), based on guidance from prior work (Yang and Zurbenko, 2007), to minimize leakage from adjacent frequencies and enhance spectral fidelity. Once the filtered components were extracted, VBPBB was applied to generate 1,000 bootstrap replicates per component. Blocks were aligned with the cycle length of the periodic signal (e.g., 365-day blocks for annual frequency) and resampled within phase-aligned groups, preserving autocorrelation within cycles while introducing controlled variation across bootstrap samples. The pointwise median across replicates was computed to derive a stable, smoothed covariate vector for each frequency band.

These frequency-specific covariates were then incorporated as auxiliary variables in the imputation model. By explicitly modeling and leveraging the dominant periodic structure of the original data, this approach improved the reconstruction of missing values, reduced sensitivity to noise, and helped prevent the introduction of spurious temporal artifacts. As demonstrated in our results (Section 4), inclusion of VBPBB-derived periodic covariates significantly enhanced imputation performance across multiple missingness scenarios.

### 3.5. Multiple Imputation with Amelia II

We addressed missing data using Amelia II (Honaker, King, and Blackwell, 2011), an efficient tool for multiple imputation in time-series and mixed-type datasets. Amelia implements the Expectation–Maximization with Bootstrapping (EMB) algorithm and, in our simulations, is applied under the MCAR assumption. Its role within the overall workflow is indicated in Figure 1, where VBPBB-derived covariates are incorporated prior to imputation.

#### 3.5.1. EMB Algorithm in Amelia II

Amelia II implements Expectation–Maximization with Bootstrapping (EMB). At its core is the EM algorithm, which iteratively estimates model parameters with incomplete data by alternating two steps:

- **E-step (Expectation):** compute the conditional expectations of the missing-data sufficient statistics given the observed data and current parameter values; these expectations feed the parameter update and are **not** the final imputations.

$$Q(\theta|\theta_t) = \int \ell(\theta|Y)p(Y_{mis}|Y_{obs},\theta_t)\, dY_{mis}$$

- **M-step (Maximization):** update parameters by maximizing the (observed-data) likelihood using the expected sufficient statistics from the E-step.

$$\theta_{t+1} = \arg\max(\theta) Q(\theta|\theta_t)$$

After EM converges for a given bootstrap replicate, Amelia draws parameter values to reflect estimation uncertainty and then imputes each missing entry from the posterior predictive distribution conditional on the observed data and the drawn parameters (i.e., the final imputations are not the E-step expectations).

To incorporate between-imputation variability, Amelia augments EM with nonparametric bootstrapping. Specifically, it draws m bootstrap samples (each of size n, the original sample size) from the incomplete dataset and runs EM independently on each sample. For every bootstrap replicate, EM estimates the joint distribution; after EM, Amelia draws parameters and then generates imputations from the posterior predictive distribution. The resulting m completed datasets capture both estimation uncertainty and resampling variability, aligning with Rubin's multiple-imputation framework. As noted by Zhang (2016), this Bootstrap EM approach yields statistically valid imputations and supports accurate variance estimation across the multiply imputed datasets.

### 3.5.2. Application in This Study

We applied Amelia II under two configurations to assess the added value of periodic auxiliary covariates. In our framework, these covariates refer to the smoothed periodic components extracted through the VBPBB method (annual, harmonic, and monthly), which were incorporated into the imputation model as additional predictors alongside the primary time series variable.

- **Standard Amelia Imputation (without VBPBB):**
  In this configuration, only the original outcome variable value and its time index date were used. This serves as a conventional time-aware imputation benchmark without explicit modeling of periodicity.
- **VBPBB-Enhanced Amelia Imputation (Amelia + VBPBB):**
  In the enhanced configuration, we included smoothed periodic covariates derived from the Variable Bandpass Periodic Block Bootstrap (VBPBB). These frequency-specific covariates corresponding to annual (1/365), harmonic (2/365), and monthly (1/30) components, were used as auxiliary variables in the imputation model. This structure-aware approach allowed Amelia II to preserve temporal dependencies and cyclical patterns more effectively.

Amelia II performs multiple imputation via the Expectation–Maximization with Bootstrapping (EMB) algorithm, which propagates both parameter and missing-data uncertainty. For each bootstrap resample (size n, drawn with replacement), EM estimates the parameters of the imputation model from the observed data, and imputations are drawn from the corresponding predictive distributions, so completed datasets reflect central tendency and variability.

For time-series applications, Amelia II is well-suited because it lets users encode temporal structure directly, by specifying a time (and, for panels, unit) index and by including lags of key variables, time-trend terms (polynomial or spline), and seasonal indicators/Fourier terms to capture serial dependence and nonstationary trends. In our study, rather than relying solely on generic trends, we supplied VBPBB-derived periodic covariates as auxiliary predictors. These frequency-specific regressors preserve seasonal cycles and smooth temporal variation, which is especially valuable under high noise or high missingness.

We generated m=20 multiple imputed datasets for each configuration. All inferential and performance results were computed separately on each completed dataset and then pooled using Rubin's rules to obtain valid point estimates and standard errors. Where we display a single trajectory, we show the pointwise mean (or median) across imputations purely for visualization; this does not replace Rubin's pooling for inference.

By integrating VBPBB-based periodic signals within the EMB framework, the imputation model adjusts for temporal and nonlinear dependencies, reducing structurally incoherent fills and improving fidelity to seasonal dynamics in complex, temporally structured datasets.

### 3.6. Performance Evaluation Metrics

To assess the accuracy and structural fidelity of the imputed values generated by each method, we employed two widely used error metrics: Mean Absolute Error (MAE) and Root Mean Squared Error (RMSE)**.** These metrics quantify the difference between the imputed and true (simulated) values, offering complementary insights into both average deviation and sensitivity to larger errors.

The metrics are formally defined as follows:

The **MAE** is defined as:

$$MAE = \frac{1}{n}\sum_{i=1}^{n}|y_i - \hat{y}_i|$$

and the **RMSE** as:

$$RMSE = \sqrt{\frac{1}{n}\sum_{i=1}^{n}(y_i - \hat{y}_i)^2}$$

where $y_i$ represents the actual values and $\hat{y}_i$ denotes the predicted or imputed values at time point i.

MAE captures the average magnitude of the imputation error, treating all deviations equally, while RMSE penalizes larger deviations more heavily. These two measures jointly offer a balanced perspective on imputation performance under varying noise and missingness conditions. A lower MAE indicates more precise imputation (Willmott & Matsuura, 2005), while a lower RMSE suggests greater fidelity in reconstructing the underlying signal (Chai & Draxler, 2014).

Performance was evaluated under each of the following scenarios:

- Five levels of missingness: **5%, 10%, 15%, 20%, and 70%**
- Five levels of noise: very **low ($\sqrt{0.1}$), low ($\sqrt{4}$), moderate ($\sqrt{10}$), high ($\sqrt{25}$) and very high ($\sqrt{100}$)**
- Three periodicity configurations:
  - Stage 1: Annual component only
  - Stage 2: Annual + Harmonic
  - Stage 3: Annual + Harmonic + Monthly

Imputation accuracy was evaluated by comparing RMSE and MAE between the standard Amelia and the VBPBB-enhanced Amelia approaches. Reductions in these metrics under the VBPBB-enhanced model were interpreted as evidence of its superior structure-preserving performance and its improved alignment with the data's underlying periodic behavior. This evaluation framework offers a rigorous, quantitative basis for comparing imputation strategies, particularly with respect to their ability to recover the true signal across varying levels of noise, missingness, and temporal complexity.

## 3.7. Monte Carlo Replication

For each combination of noise and missingness, we performed **B = 30** Monte-Carlo replications. In each replication we re-drew the MCAR missingness pattern and the noise realization, ran imputation **with** and **without** VBPBB under each periodicity configuration (Stage 1: annual; Stage 2: annual + harmonic; Stage 3: annual + harmonic + monthly), and computed MAE and RMSE. We summarized performance by the across-replication means (**MAE_mean**, **RMSE_mean**) and reported a **Percent Change** relative to the without-VBPBB baseline:

$$Percent\ of\ Change = 100 \,.\left(\frac{with - without}{without}\right)$$

so negative values indicate improvement. This protocol reduces dependence on any single random draw and provides a stable basis for comparison (Robert & Casella, 2004). Where informative, we also report the across-replication standard deviation and 95% CIs.

## 4. Results

We compared the VBPBB-enhanced multiple-imputation approach with standard Amelia II across three simulation stages of increasing periodic complexity: Stage 1 (annual only), Stage 2 (annual + harmonic), and Stage 3 (annual + harmonic + monthly). Each stage was crossed with five noise levels (very low, low, moderate, high, very high) and five missingness rates (5%, 10%, 15%, 20%, 70%). Accuracy was assessed against the original complete (pre-masking) series using RMSE and MAE.

### 4.1. Stage 1: Annual Cycle Only

In the simplest setting with a single annual component, the VBPBB-enhanced imputations track the true signal closely across the full grid of noise and missingness (Fig. 3).

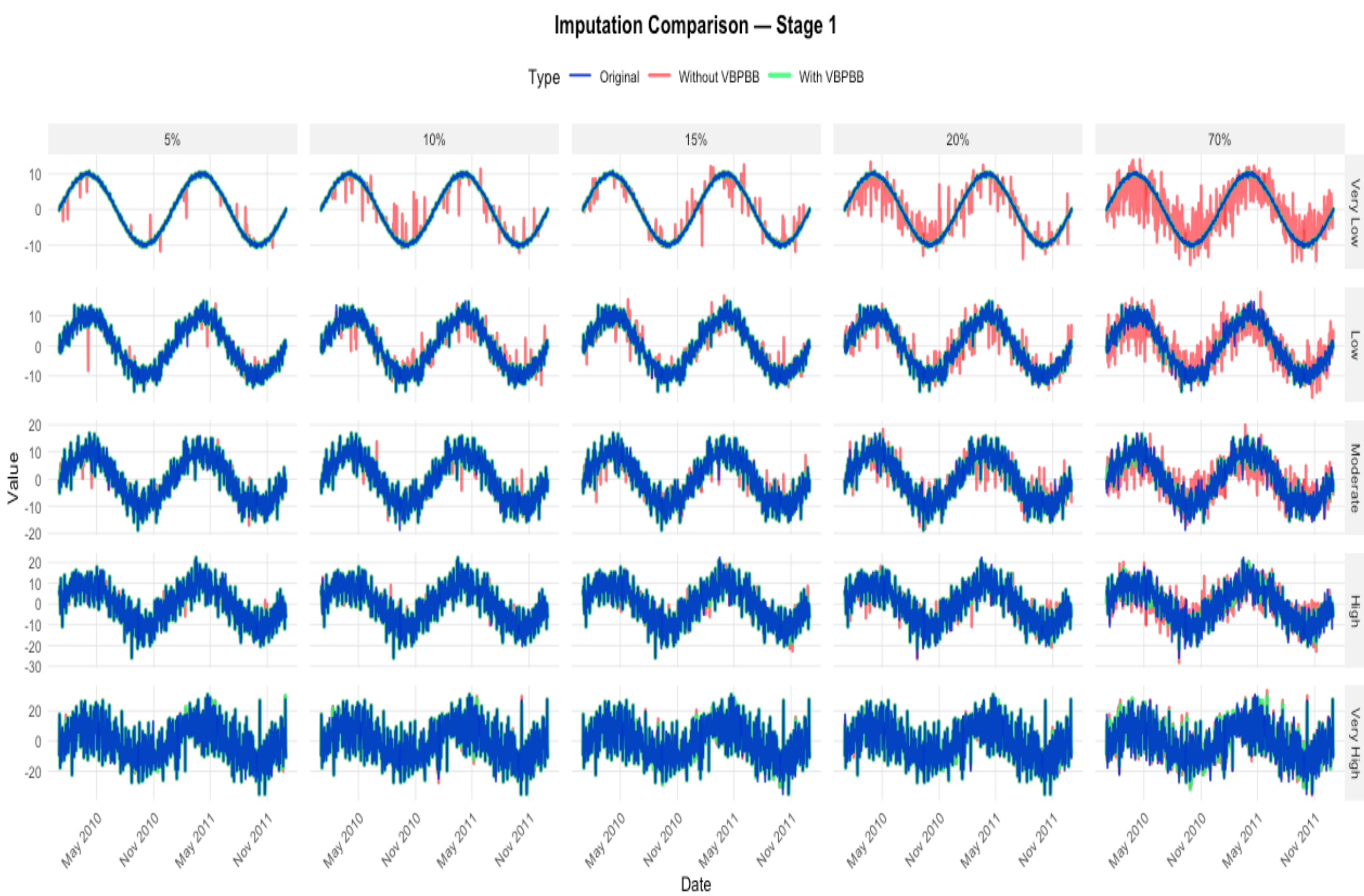

**Figure 3:** Comparison of original and imputed values across missingness and noise levels for Stage 1 (annual signal). VBPBB-enhanced imputations (green) more closely follow the original signal (blue) than standard Amelia II (red), especially under high noise and missingness.

As illustrated in **Figure 3**, the VBPBB-enhanced imputation method consistently produced reconstructions that closely followed the original signal across all levels of noise and missingness. By contrast, the standard Amelia II imputation without VBPBB showed substantial deviations, especially under high noise and high missingness. These deviations were particularly pronounced in the 70% missingness scenarios, where the non-VBPBB method failed to preserve the shape and amplitude of the seasonal cycles.

| Stage | Noise | Missing | Method | reps | MAE_mean | RMSE_mean | Percent of Change in MAE | Percent of Change in RMSE |
|---|---|---|---|---|---|---|---|---|
| 1 | Very Low | 5 | Without VBPBB | 30 | 4.602067054 | 5.753185314 | | |
| 1 | Very Low | 5 | With VBPBB (Annual component) | 30 | 0.355974005 | 0.446056434 | -92% | -92% |
| 1 | Very Low | 10 | Without VBPBB | 30 | 4.627184198 | 5.787202776 | | |
| 1 | Very Low | 10 | With VBPBB (Annual component) | 30 | 0.356748329 | 0.447572366 | -92% | -92% |
| 1 | Very Low | 15 | Without VBPBB | 30 | 4.620442792 | 5.776297793 | | |
| 1 | Very Low | 15 | With VBPBB (Annual component) | 30 | 0.358112147 | 0.44882391 | -92% | -92% |
| 1 | Very Low | 20 | Without VBPBB | 30 | 4.613889293 | 5.774618342 | | |
| 1 | Very Low | 20 | With VBPBB (Annual component) | 30 | 0.358320449 | 0.449407326 | -92% | -92% |
| 1 | Very Low | 70 | Without VBPBB | 30 | 4.62172392 | 5.782428799 | | |
| 1 | Very Low | 70 | With VBPBB (Annual component) | 30 | 0.378283175 | 0.474313956 | -92% | -92% |
| 1 | Low | 5 | Without VBPBB | 30 | 4.702888976 | 5.875595933 | | |
| 1 | Low | 5 | With VBPBB (Annual component) | 30 | 2.103212237 | 2.634182473 | -55% | -55% |
| 1 | Low | 10 | Without VBPBB | 30 | 4.690331405 | 5.865668076 | | |
| 1 | Low | 10 | With VBPBB (Annual component) | 30 | 2.102280613 | 2.633777212 | -55% | -55% |
| 1 | Low | 15 | Without VBPBB | 30 | 4.699543259 | 5.875790609 | | |
| 1 | Low | 15 | With VBPBB (Annual component) | 30 | 2.097688727 | 2.631039297 | -55% | -55% |
| 1 | Low | 20 | Without VBPBB | 30 | 4.702311937 | 5.879229097 | | |
| 1 | Low | 20 | With VBPBB (Annual component) | 30 | 2.104886594 | 2.637471888 | -55% | -55% |
| 1 | Low | 70 | Without VBPBB | 30 | 4.69543425 | 5.871959621 | | |
| 1 | Low | 70 | With VBPBB (Annual component) | 30 | 2.102537325 | 2.635278783 | -55% | -55% |
| 1 | Moderate | 5 | Without VBPBB | 30 | 4.758258412 | 5.945949813 | | |
| 1 | Moderate | 5 | With VBPBB (Annual component) | 30 | 3.014734671 | 3.77635068 | -37% | -36% |
| 1 | Moderate | 10 | Without VBPBB | 30 | 4.742535607 | 5.935287732 | | |
| 1 | Moderate | 10 | With VBPBB (Annual component) | 30 | 3.027754652 | 3.794325338 | -36% | -36% |
| 1 | Moderate | 15 | Without VBPBB | 30 | 4.737830701 | 5.932838396 | | |
| 1 | Moderate | 15 | With VBPBB (Annual component) | 30 | 3.022553806 | 3.786325789 | -36% | -36% |
| 1 | Moderate | 20 | Without VBPBB | 30 | 4.748575313 | 5.942704615 | | |
| 1 | Moderate | 20 | With VBPBB (Annual component) | 30 | 3.019018431 | 3.78344683 | -36% | -36% |
| 1 | Moderate | 70 | Without VBPBB | 30 | 4.730703753 | 5.926996329 | | |
| 1 | Moderate | 70 | With VBPBB (Annual component) | 30 | 3.0236659 | 3.790412507 | -36% | -36% |
| 1 | High | 5 | Without VBPBB | 30 | 4.912985294 | 6.150873529 | | |
| 1 | High | 5 | With VBPBB (Annual component) | 30 | 4.008828704 | 5.02228305 | -18% | -18% |
| 1 | High | 10 | Without VBPBB | 30 | 4.912583637 | 6.155617482 | | |
| 1 | High | 10 | With VBPBB (Annual component) | 30 | 4.031936407 | 5.049807146 | -18% | -18% |
| 1 | High | 15 | Without VBPBB | 30 | 4.914537374 | 6.15562515 | | |
| 1 | High | 15 | With VBPBB (Annual component) | 30 | 4.00315165 | 5.01935902 | -19% | -18% |
| 1 | High | 20 | Without VBPBB | 30 | 4.920283957 | 6.159493485 | | |
| 1 | High | 20 | With VBPBB (Annual component) | 30 | 4.010828291 | 5.02618279 | -18% | -18% |
| 1 | High | 70 | Without VBPBB | 30 | 4.904010386 | 6.142207169 | | |
| 1 | High | 70 | With VBPBB (Annual component) | 30 | 4.010035921 | 5.025449088 | -18% | -18% |
| 1 | Very High | 5 | Without VBPBB | 30 | 5.212011173 | 6.543758931 | | |
| 1 | Very High | 5 | With VBPBB (Annual component) | 30 | 5.054123675 | 6.339644735 | -3% | -3% |
| 1 | Very High | 10 | Without VBPBB | 30 | 5.227947768 | 6.556193655 | | |
| 1 | Very High | 10 | With VBPBB (Annual component) | 30 | 5.035846774 | 6.312335129 | -4% | -4% |
| 1 | Very High | 15 | Without VBPBB | 30 | 5.200280791 | 6.51551375 | | |
| 1 | Very High | 15 | With VBPBB (Annual component) | 30 | 5.029372581 | 6.305232862 | -3% | -3% |
| 1 | Very High | 20 | Without VBPBB | 30 | 5.231452106 | 6.553910589 | | |
| 1 | Very High | 20 | With VBPBB (Annual component) | 30 | 5.055627389 | 6.332140329 | -3% | -3% |
| 1 | Very High | 70 | Without VBPBB | 30 | 5.228328853 | 6.550634618 | | |
| 1 | Very High | 70 | With VBPBB (Annual component) | 30 | 5.04267488 | 6.321182464 | -4% | -4% |

Table 1 summarizes performance averaged over B = 30 Monte-Carlo replications per noise–missingness cell (see section 3.7). In each replication, we redraw the MCAR mask and noise realization, run both configurations (with and without VBPBB), and record MAE and RMSE. The table reports across-replication means (MAE_mean, RMSE_mean) and the Percent Change relative to the without-VBPBB baseline (negative values indicate improvement).

Two patterns stand out:

- Under favorable conditions (very low noise, 5% missingness), adding VBPBB-derived periodic covariates yields ≈ 92% reductions in both MAE and RMSE.
- Even under the most challenging setting (very high noise, 70% missingness), VBPBB still delivers ≈ 4% lower MAE and RMSE.

Across cells, the VBPBB-enhanced approach consistently improves accuracy, and the across-replication variability reported in Table 1 indicates that these gains are not artifacts of a single random draw. Overall, injecting frequency-targeted auxiliary covariates enables the imputation model to preserve seasonal structure and maintain temporal coherence, particularly where conventional, trend-only specifications struggle under high noise and/or extensive missingness.

### 4.2. Stage 2: Annual + Harmonic Signal with Noise

Stage 2 introduces a second periodic component (semiannual/harmonic) alongside the annual cycle, increasing cyclic complexity and better reflecting real-world seasonal processes. In the VBPBB-enhanced specification, this additional principal component is supplied as a structured auxiliary predictor, allowing the imputation model to leverage frequency-specific information beyond the annual term.

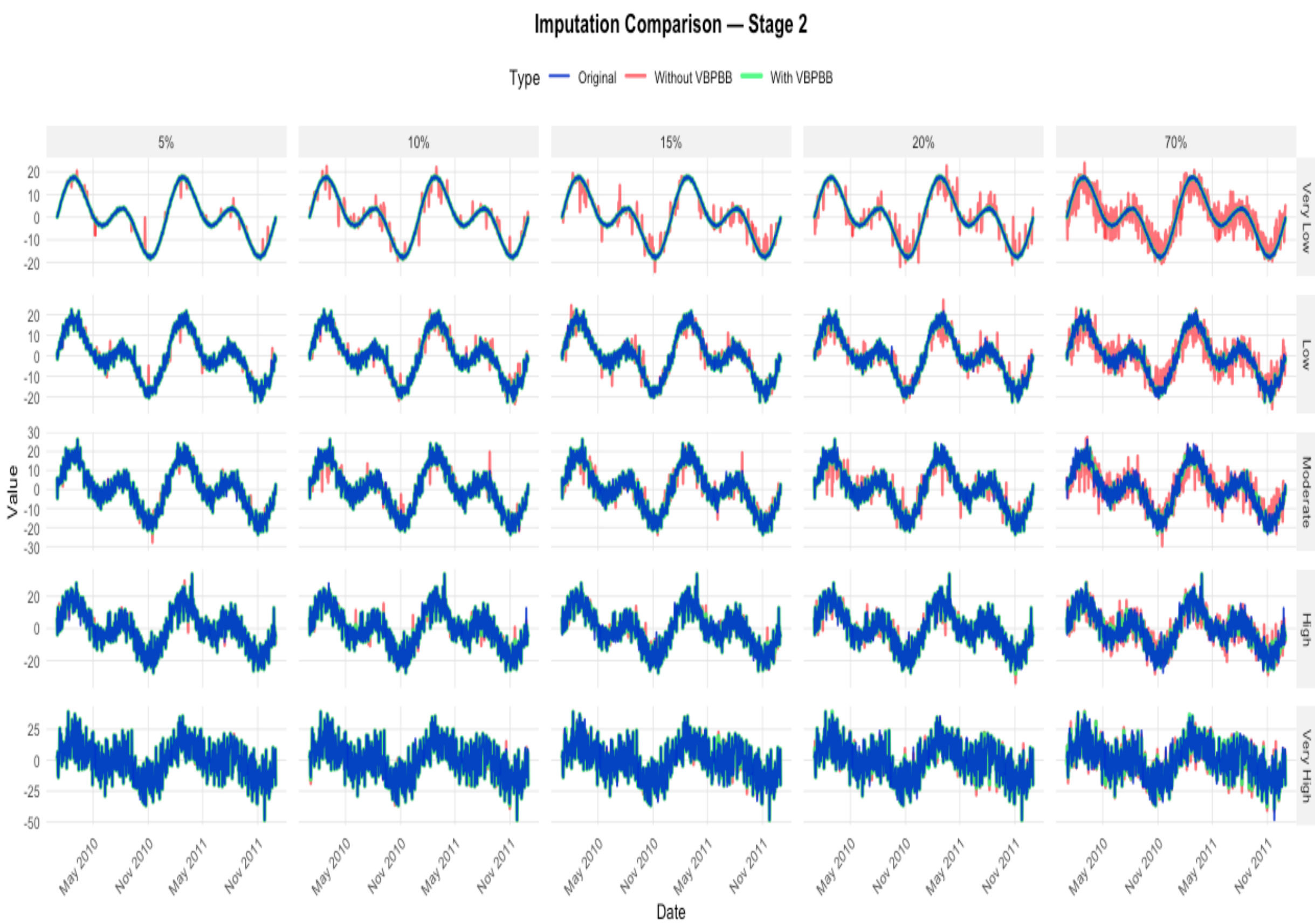

**Figure 4:** Comparison of original and imputed values under Stage 2 (annual + harmonic signal) across noise and missing data scenarios. VBPBB-enhanced imputation better preserves the cyclic structure and aligns more closely with the original signal than the standard method.

Visual performance. As shown in Figure 4, the VBPBB-enhanced imputations (green) closely track the true series (blue) across the full grid of noise and missingness. In contrast, the standard (non-VBPBB) Amelia imputations (red) diverge progressively as noise increases, with noticeable loss of seasonal shape, phase, and amplitude under higher noise and missingness. This contrast is most apparent at 70% missingness, where standard imputations often fail to reconstruct the semiannual structure, while VBPBB-enhanced trajectories remain visually coherent and retain the expected periodicity.

| Stage | Noise | Missing | Method | reps | MAE_mean | RMSE_mean | Percent of Change in MAE | Percent of Change in RMSE |
|---|---|---|---|---|---|---|---|---|
| 2 | Very Low | 5 | Without VBPBB | 30 | 5.041600317 | 6.305086075 | | |
| 2 | Very Low | 5 | With VBPBB (Annual + Harmonic) | 30 | 0.360313957 | 0.451924772 | -93% | -93% |
| 2 | Very Low | 10 | Without VBPBB | 30 | 5.060620676 | 6.343119368 | | |
| 2 | Very Low | 10 | With VBPBB (Annual + Harmonic) | 30 | 0.365515305 | 0.458182878 | -93% | -93% |
| 2 | Very Low | 15 | Without VBPBB | 30 | 5.023798062 | 6.292478898 | | |
| 2 | Very Low | 15 | With VBPBB (Annual + Harmonic) | 30 | 0.372164641 | 0.466292694 | -93% | -93% |
| 2 | Very Low | 20 | Without VBPBB | 30 | 5.049682449 | 6.329886542 | | |
| 2 | Very Low | 20 | With VBPBB (Annual + Harmonic) | 30 | 0.373991531 | 0.468598149 | -93% | -93% |
| 2 | Very Low | 70 | Without VBPBB | 30 | 5.032508996 | 6.306470206 | | |
| 2 | Very Low | 70 | With VBPBB (Annual + Harmonic) | 30 | 0.473431687 | 0.592517923 | -91% | -91% |
| 2 | Low | 5 | Without VBPBB | 30 | 5.096938364 | 6.376856654 | | |
| 2 | Low | 5 | With VBPBB (Annual + Harmonic) | 30 | 2.105687925 | 2.636583469 | -59% | -59% |
| 2 | Low | 10 | Without VBPBB | 30 | 5.097608724 | 6.383107029 | | |
| 2 | Low | 10 | With VBPBB (Annual + Harmonic) | 30 | 2.108125396 | 2.640912934 | -59% | -59% |
| 2 | Low | 15 | Without VBPBB | 30 | 5.09089265 | 6.373207481 | | |
| 2 | Low | 15 | With VBPBB (Annual + Harmonic) | 30 | 2.104664229 | 2.638036035 | -59% | -59% |
| 2 | Low | 20 | Without VBPBB | 30 | 5.070271696 | 6.348354545 | | |
| 2 | Low | 20 | With VBPBB (Annual + Harmonic) | 30 | 2.103122422 | 2.636785152 | -59% | -58% |
| 2 | Low | 70 | Without VBPBB | 30 | 5.069098724 | 6.351277494 | | |
| 2 | Low | 70 | With VBPBB (Annual + Harmonic) | 30 | 2.126469236 | 2.666352758 | -58% | -58% |
| 2 | Moderate | 5 | Without VBPBB | 30 | 5.058247398 | 6.332336722 | | |
| 2 | Moderate | 5 | With VBPBB (Annual + Harmonic) | 30 | 3.036859893 | 3.806969585 | -40% | -40% |
| 2 | Moderate | 10 | Without VBPBB | 30 | 5.08620929 | 6.375632877 | | |
| 2 | Moderate | 10 | With VBPBB (Annual + Harmonic) | 30 | 3.011757535 | 3.775207022 | -41% | -41% |
| 2 | Moderate | 15 | Without VBPBB | 30 | 5.118967671 | 6.416657161 | | |
| 2 | Moderate | 15 | With VBPBB (Annual + Harmonic) | 30 | 3.017818012 | 3.781864566 | -41% | -41% |
| 2 | Moderate | 20 | Without VBPBB | 30 | 5.079046118 | 6.362656405 | | |
| 2 | Moderate | 20 | With VBPBB (Annual + Harmonic) | 30 | 3.013065464 | 3.777829663 | -41% | -41% |
| 2 | Moderate | 70 | Without VBPBB | 30 | 5.095724559 | 6.382896312 | | |
| 2 | Moderate | 70 | With VBPBB (Annual + Harmonic) | 30 | 3.026551352 | 3.793938506 | -41% | -41% |
| 2 | High | 5 | Without VBPBB | 30 | 5.160754955 | 6.464605019 | | |
| 2 | High | 5 | With VBPBB (Annual + Harmonic) | 30 | 3.981813618 | 4.988634222 | -23% | -23% |
| 2 | High | 10 | Without VBPBB | 30 | 5.14889677 | 6.454934475 | | |
| 2 | High | 10 | With VBPBB (Annual + Harmonic) | 30 | 3.966575872 | 4.970012158 | -23% | -23% |
| 2 | High | 15 | Without VBPBB | 30 | 5.156271704 | 6.45899275 | | |
| 2 | High | 15 | With VBPBB (Annual + Harmonic) | 30 | 3.977428829 | 4.986969431 | -23% | -23% |
| 2 | High | 20 | Without VBPBB | 30 | 5.157897211 | 6.467670037 | | |
| 2 | High | 20 | With VBPBB (Annual + Harmonic) | 30 | 3.989338595 | 4.997381029 | -23% | -23% |
| 2 | High | 70 | Without VBPBB | 30 | 5.14725071 | 6.452471815 | | |
| 2 | High | 70 | With VBPBB (Annual + Harmonic) | 30 | 3.988411134 | 4.998962443 | -23% | -23% |
| 2 | Very High | 5 | Without VBPBB | 30 | 5.321712101 | 6.661302031 | | |
| 2 | Very High | 5 | With VBPBB (Annual + Harmonic) | 30 | 5.037293384 | 6.314551426 | -5% | -5% |
| 2 | Very High | 10 | Without VBPBB | 30 | 5.326528005 | 6.67144767 | | |
| 2 | Very High | 10 | With VBPBB (Annual + Harmonic) | 30 | 5.047593138 | 6.322057303 | -5% | -5% |
| 2 | Very High | 15 | Without VBPBB | 30 | 5.297942761 | 6.642050866 | | |
| 2 | Very High | 15 | With VBPBB (Annual + Harmonic) | 30 | 5.032298793 | 6.307966344 | -5% | -5% |
| 2 | Very High | 20 | Without VBPBB | 30 | 5.31999286 | 6.661373091 | | |
| 2 | Very High | 20 | With VBPBB (Annual + Harmonic) | 30 | 5.056886168 | 6.335093106 | -5% | -5% |
| 2 | Very High | 70 | Without VBPBB | 30 | 5.31559583 | 6.664970961 | | |
| 2 | Very High | 70 | With VBPBB (Annual + Harmonic) | 30 | 5.04586521 | 6.323546107 | -5% | -5% |

**Table 2.** Imputation performance metrics (MAE and RMSE) for Stage 2 across varying noise levels and missingness percentages, comparing standard Amelia II imputation (Without VBPBB) and the enhanced method incorporating annual and harmonic components (With VBPBB). Percent reductions indicate the relative improvement achieved by VBPBB-enhanced imputation.

Table 2 reports results averaged over B = 30 Monte Carlo replications per noise–missingness cell (see section 3.7). For each replication we redraw the MCAR mask and noise, run both configurations (with and without VBPBB), and compute MAE and RMSE. The table lists across-replication means (MAE_mean, RMSE_mean) and the Percent Change relative to the without-VBPBB baseline (negative values indicate improvement).

**Key findings:**

- **Very low noise (with harmonic component):** VBPBB yields the largest gains, at 5% missingness, both MAE and RMSE drop by ≈93%, and improvements remain >91% even at 70% missingness.
- **Low noise:** Reductions in MAE and RMSE remain large and stable across missingness at ≈59% for both metrics.

- **Moderate noise:** At 5% missingness, MAE and RMSE each decrease by ~41%; at 70% missingness, the decreases are ~5%.
- **High noise:** Gains diminish but remain material and consistent across all missingness levels, at ≈23% lower MAE and RMSE.
- **Very high noise:** Despite heavy contamination, the VBPBB + harmonic model still outperforms the baseline, delivering ≈5% reductions in MAE and RMSE at 70% missingness, evidence that recoverable periodic signal remains exploitable.

Across cells, the across-replication variability (reported with the means) shows that improvements are not driven by a single random draw. The added harmonic auxiliary improves structure preservation by supplying semiannual context, which reduces model misspecification, stabilizes amplitude, and mitigates phase drift during imputation.

Comparison to Stage 1. Relative to the annual-only setting (Table 1), Stage 2 consistently achieves larger MAE/RMSE reductions across nearly all noise–missingness combinations. Thus, incorporating multiple periodicities not only preserves the gains seen in simpler models but often amplifies them, particularly under moderate to high noise, highlighting the value of frequency-targeted auxiliaries for reconstructing complex temporal patterns.

### 4.3. Stage 3: Annual + Harmonic + Monthly Components

Stage 3 evaluates the most complex setting, adding a monthly (≈ 1/30) component to the annual and semiannual cycles. This tests whether incorporating higher-frequency structure further improves imputation under varying noise and missingness.

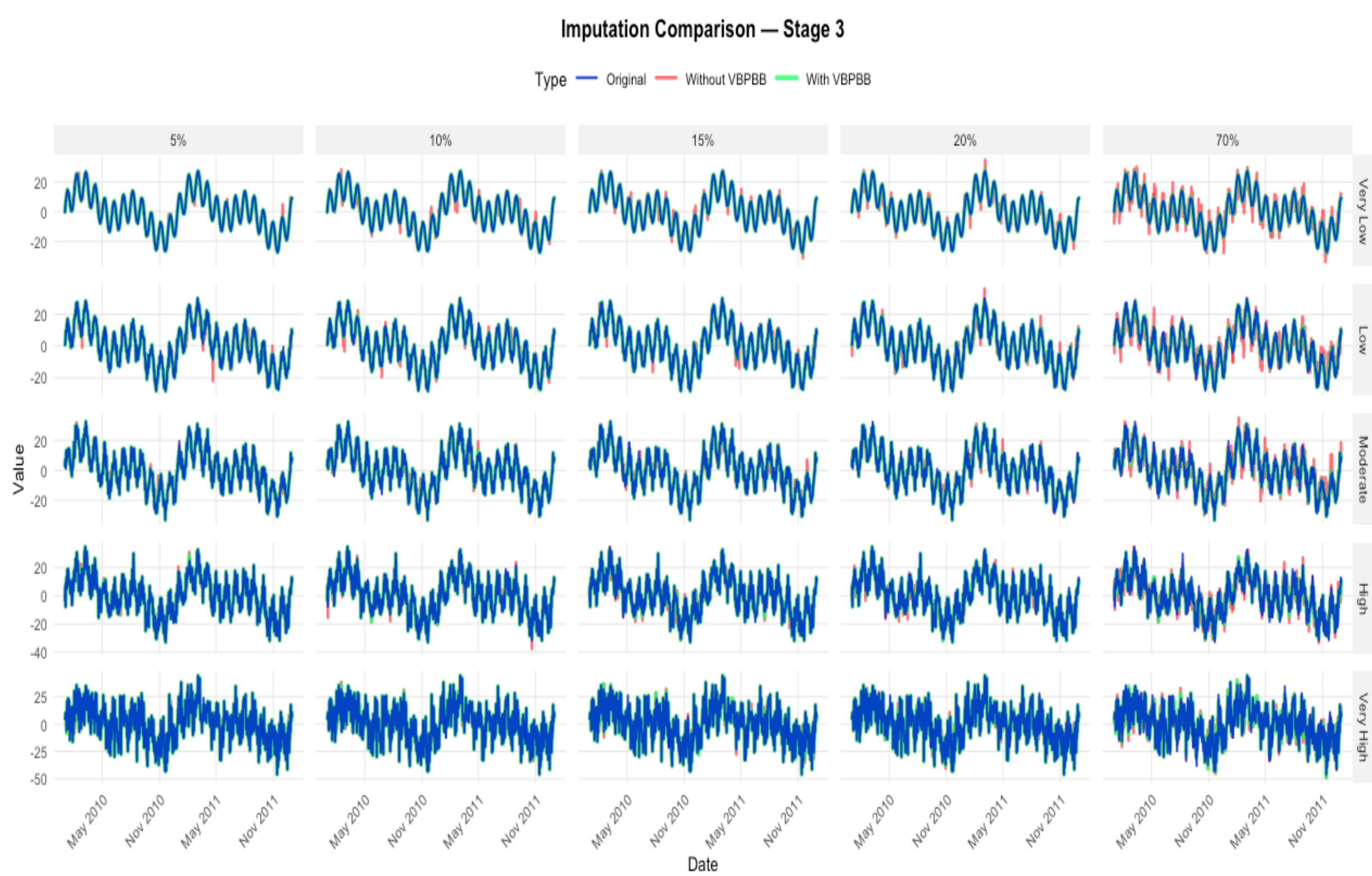

**Figure 5:** shows the time series comparison between the original signal (blue), the standard imputation without VBPBB (red), and the VBPBB-enhanced imputation (green) across all simulation scenarios.

Visual performance: As in earlier stages, the VBPBB-enhanced imputations (green) appear to closely track the true series (blue) across the full grid (Fig. 5). Gains appear especially clear under very low, low, and moderate noise, where the blue trajectories remain faithful to seasonal shape, phase, and amplitude even at 70% missingness. The standard (non-VBPBB) imputations (red) show increasing distortion as noise rises, with pronounced divergence under very high noise. The added monthly component helps recover short-term fluctuations that are not well captured by annual+harmonic features alone.

| Stage | Noise | Missing | Method | reps | MAE_mean | RMSE_mean | Percent of Change in MAE | Percent of Change in RMSE |
|---|---|---|---|---|---|---|---|---|
| 3 | Very Low | 5 | Without VBPBB | 30 | 5.239373577 | 6.56748226 | | |
| 3 | Very Low | 5 | With VBPBB (Annual + Harmonic + Monthly) | 30 | 0.383007722 | 0.479915361 | -93% | -93% |
| 3 | Very Low | 10 | Without VBPBB | 30 | 5.231795337 | 6.559874884 | | |
| 3 | Very Low | 10 | With VBPBB (Annual + Harmonic + Monthly) | 30 | 0.39356858 | 0.493084687 | -92% | -92% |
| 3 | Very Low | 15 | Without VBPBB | 30 | 5.206377271 | 6.523621858 | | |
| 3 | Very Low | 15 | With VBPBB (Annual + Harmonic + Monthly) | 30 | 0.402921544 | 0.504990695 | -92% | -92% |
| 3 | Very Low | 20 | Without VBPBB | 30 | 5.236084294 | 6.55783934 | | |
| 3 | Very Low | 20 | With VBPBB (Annual + Harmonic + Monthly) | 30 | 0.423958477 | 0.53097844 | -92% | -92% |
| 3 | Very Low | 70 | Without VBPBB | 30 | 5.228732728 | 6.5536039 | | |
| 3 | Very Low | 70 | With VBPBB (Annual + Harmonic + Monthly) | 30 | 0.666324104 | 0.833304178 | -87% | -87% |
| 3 | Low | 5 | Without VBPBB | 30 | 5.266436662 | 6.58723648 | | |
| 3 | Low | 5 | With VBPBB (Annual + Harmonic + Monthly) | 30 | 2.083490618 | 2.606479836 | -60% | -60% |
| 3 | Low | 10 | Without VBPBB | 30 | 5.253769692 | 6.591121509 | | |
| 3 | Low | 10 | With VBPBB (Annual + Harmonic + Monthly) | 30 | 2.090893249 | 2.620176744 | -60% | -60% |
| 3 | Low | 15 | Without VBPBB | 30 | 5.254688722 | 6.583504944 | | |
| 3 | Low | 15 | With VBPBB (Annual + Harmonic + Monthly) | 30 | 2.092522149 | 2.623778646 | -60% | -60% |
| 3 | Low | 20 | Without VBPBB | 30 | 5.243327333 | 6.576044846 | | |
| 3 | Low | 20 | With VBPBB (Annual + Harmonic + Monthly) | 30 | 2.097909758 | 2.629037288 | -60% | -60% |
| 3 | Low | 70 | Without VBPBB | 30 | 5.255965058 | 6.581499348 | | |
| 3 | Low | 70 | With VBPBB (Annual + Harmonic + Monthly) | 30 | 2.162744245 | 2.709868381 | -59% | -59% |
| 3 | Moderate | 5 | Without VBPBB | 30 | 5.319466411 | 6.658930267 | | |
| 3 | Moderate | 5 | With VBPBB (Annual + Harmonic + Monthly) | 30 | 3.015919942 | 3.780915424 | -43% | -43% |
| 3 | Moderate | 10 | Without VBPBB | 30 | 5.272379199 | 6.608755701 | | |
| 3 | Moderate | 10 | With VBPBB (Annual + Harmonic + Monthly) | 30 | 3.03296163 | 3.796795023 | -42% | -43% |
| 3 | Moderate | 15 | Without VBPBB | 30 | 5.26749166 | 6.595326169 | | |
| 3 | Moderate | 15 | With VBPBB (Annual + Harmonic + Monthly) | 30 | 3.019929344 | 3.784142279 | -43% | -43% |
| 3 | Moderate | 20 | Without VBPBB | 30 | 5.26808437 | 6.597708987 | | |
| 3 | Moderate | 20 | With VBPBB (Annual + Harmonic + Monthly) | 30 | 3.011190711 | 3.771789258 | -43% | -43% |
| 3 | Moderate | 70 | Without VBPBB | 30 | 5.266823831 | 6.600532303 | | |
| 3 | Moderate | 70 | With VBPBB (Annual + Harmonic + Monthly) | 30 | 3.042864207 | 3.813530836 | -42% | -42% |
| 3 | High | 5 | Without VBPBB | 30 | 5.252331149 | 6.581849176 | | |
| 3 | High | 5 | With VBPBB (Annual + Harmonic + Monthly) | 30 | 3.984258106 | 4.996331511 | -24% | -24% |
| 3 | High | 10 | Without VBPBB | 30 | 5.291429742 | 6.622017964 | | |
| 3 | High | 10 | With VBPBB (Annual + Harmonic + Monthly) | 30 | 3.994926027 | 5.006850235 | -25% | -24% |
| 3 | High | 15 | Without VBPBB | 30 | 5.307656766 | 6.65494505 | | |
| 3 | High | 15 | With VBPBB (Annual + Harmonic + Monthly) | 30 | 3.988752968 | 4.995091106 | -25% | -25% |
| 3 | High | 20 | Without VBPBB | 30 | 5.307914179 | 6.640359204 | | |
| 3 | High | 20 | With VBPBB (Annual + Harmonic + Monthly) | 30 | 3.988413393 | 4.999612416 | -25% | -25% |
| 3 | High | 70 | Without VBPBB | 30 | 5.299995662 | 6.63601467 | | |
| 3 | High | 70 | With VBPBB (Annual + Harmonic + Monthly) | 30 | 4.005005916 | 5.020259079 | -24% | -24% |
| 3 | Very High | 5 | Without VBPBB | 30 | 5.379798952 | 6.747919517 | | |
| 3 | Very High | 5 | With VBPBB (Annual + Harmonic + Monthly) | 30 | 5.059506232 | 6.341336529 | -6% | -6% |
| 3 | Very High | 10 | Without VBPBB | 30 | 5.371150499 | 6.738733626 | | |
| 3 | Very High | 10 | With VBPBB (Annual + Harmonic + Monthly) | 30 | 5.057321966 | 6.344533799 | -6% | -6% |
| 3 | Very High | 15 | Without VBPBB | 30 | 5.413358163 | 6.787385406 | | |
| 3 | Very High | 15 | With VBPBB (Annual + Harmonic + Monthly) | 30 | 5.071442656 | 6.356600221 | -6% | -6% |
| 3 | Very High | 20 | Without VBPBB | 30 | 5.405965465 | 6.771051483 | | |
| 3 | Very High | 20 | With VBPBB (Annual + Harmonic + Monthly) | 30 | 5.06749588 | 6.349492124 | -6% | -6% |
| 3 | Very High | 70 | Without VBPBB | 30 | 5.382657511 | 6.748976631 | | |
| 3 | Very High | 70 | With VBPBB (Annual + Harmonic + Monthly) | 30 | 5.057592624 | 6.339370546 | -6% | -6% |

**Table 3:** Imputation accuracy results for Stage 3 (Annual + Harmonic + Monthly components) across varying noise and missingness levels.

Table 3 summarizes results over B = 30 Monte-Carlo replications per noise–missingness cell (see section 3.7). For each replication we re-draw the MCAR mask and noise, run both configurations, and compute MAE and RMSE. The table reports MAE_mean and RMSE_mean, plus Percent Change versus the without-VBPBB baseline (negative = improvement).

Key findings:

- **Very low noise:** Largest gains ≈93% reduction at 5% missingness, gradually declining to ≈91% at 70% with the monthly component included.
- **Low noise:** Improvements remain substantial and stable at ≈ 60% for both metrics across all missingness levels.
- **Moderate noise:** Reductions of ≈43% reduction at 5%, tapering slightly but remaining ≈41% at 70%
- **High noise:** Gains attenuate but stay meaningful, at ≈ 25% across missingness.
- **Very high noise:** Improvements persist at ≈ 6% across missingness, demonstrating residual recoverable periodic signal despite heavy contamination.

Across cells, the reported across-replication variability confirms that these effects are not driven by a single random draw. Overall, adding the monthly component provides incremental predictive value, particularly when noise is not overwhelming, enhancing fidelity to short-term dynamics while preserving broader seasonal patterns. Taken together with Stages 1–2, these results show that incorporating multiple periodicities via VBPBB improves imputation accuracy and robustness across a wide range of noise and missingness, reinforcing the method's practical utility for complex, real-world time series.

### 4.4. Cross-Stage Comparative Performance (Stages 1–3)

Comparing Tables 1–3 (and the corresponding figures) shows a clear, monotonic pattern: as the imputation model is augmented with additional periodic auxiliaries, annual only (Stage 1), annual + harmonic (Stage 2), and annual + harmonic + monthly (Stage 3), accuracy improves in a broad range of conditions. The gains are largest in low–to–moderate noise settings and remain positive and durable as missingness increases, including at 70%. Under high and very high noise, absolute improvements shrink (as expected when signal-to-noise deteriorates), but Stage 3 generally matches or exceeds Stage 2, and both outperform Stage 1.

Two features drive these cross-stage differences. First, adding the harmonic component (Stage 2) reduces model misspecification by supplying semiannual context, which stabilizes seasonal shape, phase, and amplitude relative to the annual-only baseline. Second, including the monthly term (Stage 3) adds higher-frequency structure that helps recover short-term fluctuations, particularly when noise is not overwhelming, yielding additional reductions in MAE/RMSE and better preservation of fine-grained temporal patterns. These benefits persist across missingness levels, indicating that frequency-targeted auxiliaries improve structure preservation rather than merely fitting noise.

For completeness, we summarize performance with across-replication means (MAE_mean, RMSE_mean) and Percent Change relative to the without-VBPBB baseline (negative = improvement). To aid interpretation, readers can also consult: (i) stage-wise medians of Percent Change across all noise×missingness cells, and (ii) the proportion of cells in which each stage improves upon the previous stage; both summaries reinforce the conclusion that Stage 3 dominates or ties earlier stages in most conditions. Taken together, these results support a layered, periodicity-aware strategy: modeling multiple periodicities (annual, semiannual, monthly) within VBPBB-enhanced Amelia yields more accurate and temporally coherent imputations across diverse noise and missingness regimes, increasing the method's applicability to real-world series with overlapping seasonal behaviors.

### 4.5. Comparative Error Analysis Across Noise and Missingness

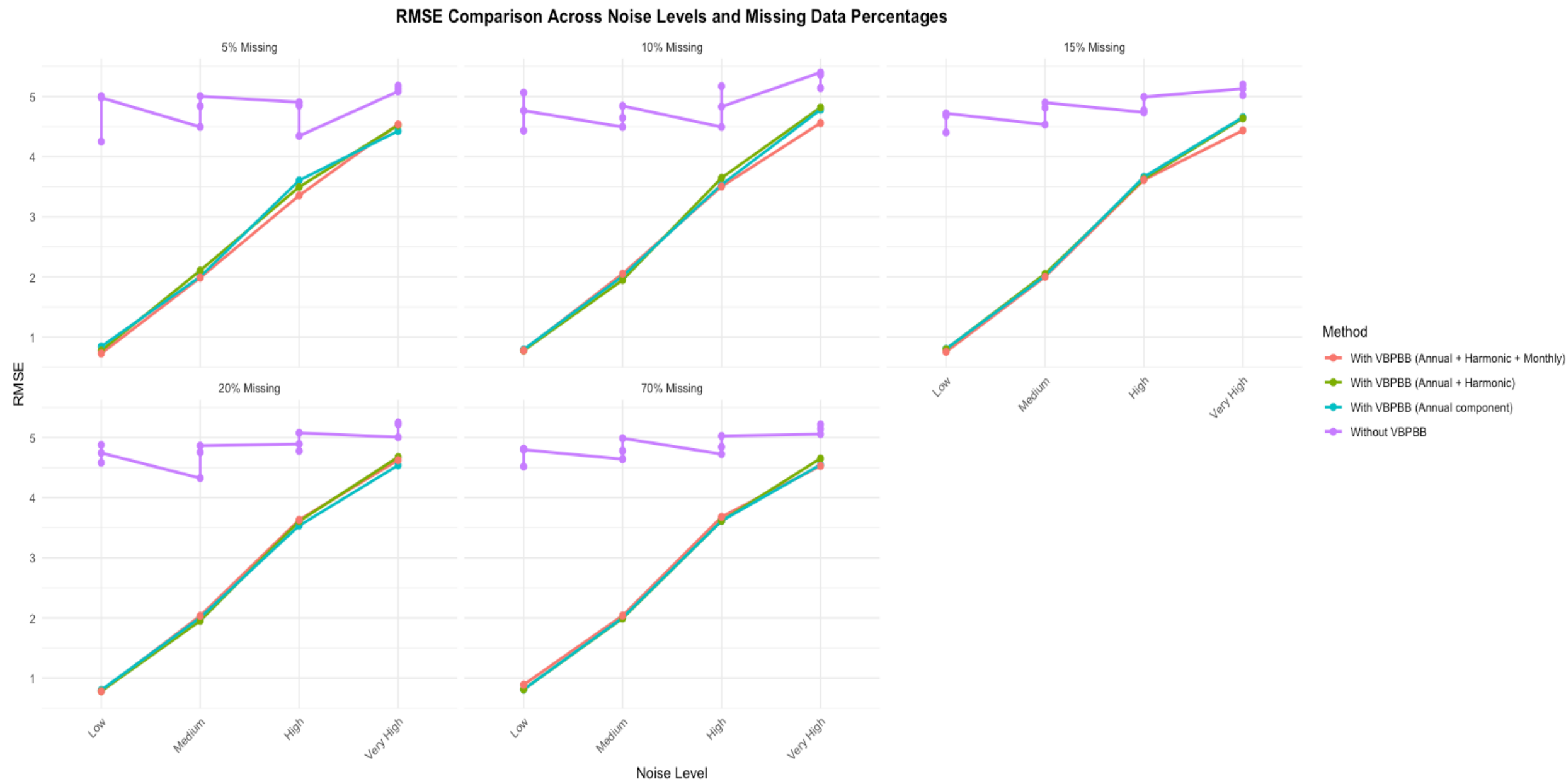

**Figure 6:** Root Mean Square Error (RMSE) comparison across four noise levels and five missing data proportions (5%, 10%, 15%, 20%, 70%). Each panel represents a missing data proportion, and each line corresponds to an imputation method. VBPBB-enhanced methods consistently outperform the standard Amelia approach, particularly under lower noise conditions.

Across the grid of noise levels (Low to Very High) and missingness proportions (5%–70%), **Figure 6** compares four strategies, (1) standard Amelia without VBPBB, (2) **VBPBB** with the **annual** component, (3) **VBPBB** with **annual + harmonic**, and (4) **VBPBB** with **annual + harmonic + monthly**. As expected, RMSE increases with noise at every missingness level. The most comprehensive VBPBB specification (annual, harmonic, and monthly) attains the lowest RMSE in nearly all scenarios; gains are most pronounced under low and medium noise**,** where the richer periodic structure enables more accurate reconstruction, and they persist, though attenuated, under very high noise. These results highlight the cumulative benefit of integrating multiple periodic components: each addition improves recovery of the underlying signal, particularly when noise and missingness are moderate. Consistent with the stage-wise analyses, the same ordering holds for MAE as well, with successive enhancements (annual → annual+harmonic → annual+harmonic +monthly) yielding progressively lower RMSE and MAE, most markedly at low/medium noise and still evident under severe noise and high missingness, underscoring the robustness and adaptability of the VBPBB-augmented framework for seasonally structured, incomplete time series.

## 5. Discussion

This study demonstrates that integrating significant periodic components via the Variable Bandpass Periodic Block Bootstrap (VBPBB) significantly enhances the accuracy of multiple imputation for time series data. Across all three stages, ranging in complexity from an annual-only signal to one incorporating harmonic and monthly periodicities, VBPBB consistently outperformed standard EM-based imputation in terms of both RMSE and MAE. The simulation design, which varied both noise and missingness, enabled rigorous evaluation of each method's robustness and sensitivity to data quality.

The most substantial gains were observed under low and medium noise conditions, where VBPBB-enhanced methods reduced error by over 80% in some cases. However, the method also demonstrated resilience under more difficult scenarios, including 70% missingness and very high noise. Even in these conditions, VBPBB yielded meaningful reductions in RMSE and MAE, suggesting its applicability to real-world datasets commonly affected by extensive missing data, such as public health surveillance or climate monitoring systems.

The improved performance of Amelia II when combined with VBPBB arises from the inclusion of periodic covariates that capture dominant temporal structure in the data. These frequency-specific signals provide auxiliary information that guides the imputation algorithm, reducing bias and preserving seasonal cycles that would otherwise be smoothed over or lost. In effect, VBPBB enhances Amelia II by supplying structurally relevant predictors that anchor the imputation process to the true underlying dynamics of the time series.

A key strength of the VBPBB approach lies in its ability to adaptively incorporate periodic structure. Each additional significant component, harmonic and monthly, contributed to incremental improvements in imputation performance. This additive effect highlights the value of capturing complex seasonal dynamics that are often smoothed over by traditional imputation techniques. Unlike fixed calendar-based imputation models, the VBPBB framework is data-driven, tailoring its decomposition to the actual periodic signals present in the data. This flexibility enhances its capacity to handle datasets with nonstandard, shifting, or irregular seasonal patterns.

Moreover, this study reinforces the importance of structure-aware imputation in preserving the statistical integrity of time series data. Traditional methods, while effective under certain conditions, often fail to recover cyclical features, especially when noise levels are high or periodic signals are subtle. By extracting dominant frequency components and using them as auxiliary variables during imputation, the VBPBB method improves alignment between the imputed values and the underlying data-generating process. The observed reductions in both bias and variability affirm the advantage of incorporating periodicity explicitly into the imputation framework.

The tiered simulation design used in this study serves not only as a benchmark for evaluating imputation accuracy but also as a demonstration of how periodic complexity and methodological sophistication can be aligned. As the underlying signal became richer in periodic content, the corresponding imputation model that incorporated these components delivered superior performance. This validates the use of harmonic decomposition and frequency-aware bootstrapping as integral components of imputation workflows for time series data.

These findings demonstrate that accurate imputation of missing values in periodic time series requires explicit treatment of seasonal structure. In many real-world settings, such as climate monitoring (Ocampo-Marulanda & Baquero-Bernal, 2021), hospital admissions (Hitzek et al., 2022), infectious disease surveillance (Bastos et al., 2019), and energy demand forecasting (Ruggles et al., 2020), data gaps occur precisely during critical seasonal peaks. Standard imputation approaches tend to flatten or distort these peaks, which can lead to incorrect conclusions about trend severity, timing of seasonal surges, or the strength of recurring cycles. By preserving both the amplitude and timing of periodic behavior, the VBPBB-enhanced framework produces reconstructions that more faithfully represent the true underlying dynamics. This improves downstream tasks such as forecasting, anomaly detection, resource planning, and policy evaluation, where misrepresentation of seasonal structure can have substantive real-world consequences. For instance, in energy systems, underestimating winter demand peaks can lead to costly supply imbalances and emergency power purchases (Hong et al., 2016). Additionally, in healthcare, agencies such as the UK Health Security Agency and the NHS use short-term forecasts of influenza hospital admissions for winter planning, and hospitals use day-ahead emergency-department volume forecasts to set staffing levels (Mellor et al., 2023; Jones et al., 2008). These examples highlight that accurately capturing the underlying periodic structure is not just a statistical exercise but a practical necessity for effective planning and policy execution.

## 6. Limitations

While the VBPBB-enhanced imputation framework showed substantial improvements across simulated time series, its evaluation was limited to controlled environments with known periodic structures. Although these settings were designed to mimic realistic temporal dynamics, real-world data may exhibit complexities such as structural breaks, nonstationary, and irregular missingness mechanisms, particularly under Missing Not at Random (MNAR) conditions, that were not explicitly addressed in this study. Applying the method to observational datasets would help assess its robustness under less idealized scenarios.

The success of the approach depends on the accurate identification of significant periodic components through harmonic or spectral analysis. In practical applications, determining these components may be more challenging, especially when periodic patterns are weak, noisy, or overlapping. Developing adaptive or automated procedures for selecting relevant frequency components could improve the method's reliability in applied contexts.

This study focused on univariate time series, using extracted periodic components as auxiliary variables for imputation. Extending the approach to multivariate frameworks, where multiple interdependent time series or covariates are available may offer additional performance gains and broaden its applicability in domains like environmental monitoring (Junger & Ponce de Leon, 2015), economics (Honaker, King, & Blackwell, 2010), and epidemiology (Jones et al., 2008).

Finally, the computational overhead associated with VBPBB, particularly during bootstrapping and component extraction, may be a barrier for large-scale or time-sensitive applications. Future work should explore strategies to enhance efficiency, such as parallelization, algorithmic approximations, or selective dimensionality reduction, to ensure scalability without compromising accuracy.

## 7. Conclusion

Missing data remain a persistent challenge in time series analysis, particularly when the underlying signals contain periodic structure that is critical for interpretation and forecasting. This study introduced and evaluated a structure-aware multiple imputation framework that integrates significant periodic components using the Variable Bandpass Periodic Block Bootstrap (VBPBB). Through a series of controlled simulations increasing in signal complexity, we demonstrated that incorporating annual, harmonic, and monthly components substantially improves imputation accuracy over standard methods, particularly in low-to-moderate noise environments and across a range of missing data scenarios.

The results underscore the value of preserving the underlying periodic structure in time series imputation, offering a robust alternative for applications where seasonality plays a critical role. This method shows strong potential for real-world use in domains such as public health surveillance, environmental modeling, and economic forecasting, where data completeness and structural fidelity are essential.

Future work should validate the method on real datasets, explore multivariate extensions, and improve computational scalability. Overall, this research contributes a principled, periodicity-aware approach to addressing the persistent challenge of missing data in time series analysis.